\documentclass[runningheads,a4paper]{llncs}

\usepackage{amssymb}
\usepackage{graphicx}
\usepackage[hyphens]{url}
\usepackage{hyperref}
\usepackage{xcolor}
\usepackage{breakcites}

\begin{document}

\title{Current and Emergent Economic Impacts of Covid-19 and Brexit on UK Fresh Produce and Horticultural Businesses}

\titlerunning{Economic Impacts of Covid-19 and Brexit on UK Horticulture}

\author{Lilian Korir\textsuperscript{1}
\and
Archie Drake\textsuperscript{1}
\and
Martin Collison\textsuperscript{2}
\and
Tania Carolina Camacho-Villa\textsuperscript{1}
\and
Elizabeth Sklar\textsuperscript{1,*}
\and
Simon Pearson\textsuperscript{1}}
\authorrunning{Korir, et al.}

\institute{Lincoln Institute for Agri-food Technology (LIAT), College of Science, University of Lincoln, UK,
\emph{\{lkorir,esklar,spearson\}@lincoln.ac.uk}
\and
Collison \& Associates Limited}

\maketitle

\begin{abstract}
This paper describes a study designed to investigate the current and emergent impacts of Covid-19 and Brexit on UK horticultural businesses. Various characteristics of UK horticultural production, notably labour reliance and import dependence, make it an important sector for policymakers concerned to understand the effects of these disruptive events as we move from 2020 into 2021. The study design prioritised timeliness, using a rapid survey to gather information from a relatively small ($n=19$) but indicative group of producers. The main novelty of the results is to suggest that a very substantial majority of producers either plan to scale back production in 2021 (47\%) or have been unable to make plans for 2021 because of uncertainty (37\%). The results also add to broader evidence that the sector has experienced profound labour supply challenges, with implications for labour cost and quality. The study discusses the implications of these insights from producers in terms of productivity and automation, as well as in terms of broader economic implications. Although automation is generally recognised as the long-term future for the industry (89\%), it appeared in the study as the second most referred short-term option (32\%) only after changes to labour schemes and policies (58\%). Currently, automation plays a limited role in contributing to the UK’s horticultural workforce shortage due to economic and socio-political uncertainties. The conclusion highlights policy recommendations and future investigative intentions, as well as suggesting methodological and other discussion points for the research community.
\end{abstract}

\section{Introduction}
\label{sec:intro}

UK horticultural production is experiencing significant changes as political and economic conditions shift. Whilst Covid-19 impacts dominated 2020~\cite{mitchell_impact_2020}, the closing months of the year have seen questions about Brexit-related issues around labour supply, trade and competitiveness return to the foreground. The sector is noted for its dependence on seasonal migrant labour, not only in the UK but globally.  Furthermore, UK fresh food supplies generally have a high dependence on imports which peaks in the winter and early spring~\cite{grimwood_migrant_2017,defra_results_2019}. 

The Current and Emergent Economic Impacts of Covid-19 and Brexit on UK Fresh Produce and Horticultural Businesses, especially soft fruit production, is the main reason for agriculture needing to find about 64,000 seasonal workers each year (by DEFRA's estimate)~\cite{ons_labour_2018}.
Employment is growers' single most important cost, and Covid-19 restrictions have increased that cost significantly~\cite{pelham_potential_2020}.
Meanwhile, changes in the immigration regime as part of Brexit are disrupting the supply of EU workers on whose skills and experience many producers have come to rely.

Accordingly, seasonal labour supply has become the main focus of policy in this sector, reflected in evidence to the Environment, Food and Rural Affairs Committee (EFRA's) inquiry into labour in the food supply chain~\cite{environment_food_and_rural_affairs_committee_labour_nodate}.

In its evidence to the EFRA Select Committee the government emphasised the potential for substitution from the UK's domestic labour supply, supported through schemes like`Pick for Britain'~\cite{ahdb_pick_nodate}.
Industry experience through 2020 indicates that this is only a very partial solution though, and in parallel with the publication of the EFRA Select Committee report, the government announced that the Seasonal Agricultural Workers Scheme (SAWS) for 2021 would be increased to 30,000 places compared to 10,000 in 2020~\cite{defra_reap_2020e}.
However, this is only a one year scheme and there remains considerable uncertainty about how many places will be allowed under SAWS from 2022 onwards.

Furthermore, the NFU report on the Potential Implications of Covid-19 for the Cost of Production of UK Fruits and Vegetables in 2020~\cite{pelham_potential_2020} gives a good impression of some of the issues that farmers have reported in practice: there are not as many suitable UK resident workers available as many imagine; UK residents like the idea of this work but then are much less likely than migrant workers to show up, stay for the agreed period; and, overall, UK residents are less productive workers in these roles.

What remains unclear is whether supply chain vulnerability and the potential for price increases could represent an opportunity for UK industry expansion.  However, the negatives impact of price rises has also been noted if healthy eating becomes more difficult~\cite{garnett_vulnerability_2020,seferidi_impacts_2019}.
In some circumstances labour constraints might spur innovation and consequently improved productivity~\cite{acemoglu_when_2010}. 

Around the world, the highly disruptive conditions of 2020 have prompted scientists to work more quickly; and specifically in agricultural and food systems research to consider `short-term effects as well as those that may be long-lasting or permanent'~\cite{kupferschmidt_race_2020,stephens_editorial_2020}.
This study was designed to provide timely evidence for UK policymakers drawn directly from UK horticultural producers' perspectives in late November 2020. Its principal goal was to analyse the current and emergent impacts related to Covid-19 and Brexit on UK horticultural businesses, especially domestic producers of fruit and vegetables.

\section{Methodology}
\label{sec:methodology}

This study used a rapid survey instrument to gather information from UK producers. A short questionnaire was developed, consisting principally of open-ended questions focused on three key themes:
\begin{enumerate}
\item The impact of Covid-19 and Brexit on 2020 labour availability and on businesses' 2021 production plans;
\item Changes observed in labour productivity during 2020 and associated impacts on market competitiveness; and
\item Market growth expectations and the means to address any growth in demand over the short- and long-term.
\end{enumerate}
The questionnaire was offered to an initial list of 21 participants comprising existing business contacts of the Lincoln Institute for Agri-Food Technology\footnote{University of Lincoln, UK, \url{https://www.lincoln.ac.uk/home/liat/}}, and the sample then developed using snowballing. 
Data collection was undertaken between November 19th and December 1st, offering a choice of online, email or telephone response. 
A final total of 19 producer responses were obtained. 
The information collected was mainly qualitative and subjected to thematic analysis, with frequency description for those questions generating quantitative data.  
Analysis was supported by a detailed review of secondary information from research publications, as well as policy material and social media.
The method in this study is consistent with the current global uptick in the use of rapid surveys and appraisals~\cite{eriksson_changes_2020,harris_food_2020}.
These methods aim to generate best-available information to support decisions in a timely manner, especially responding to policymakers' interest in current evidence to help reduce uncertainty~\cite{shaxson_is_2005}.

\section{Findings}
\label{sec:findings}

The analysis of this study was prefaced with reflections on strengths and weaknesses.
Total responses covered about one sixth of total domestic horticultural production by value, over 5\% of the land area used for fresh produce and horticulture, and about 15\% of total UK employment of seasonal workers.
In total, there were 19 responses. 

\subsection{Descriptive results.}
This descriptive analysis section provides an overview of the participants involved in this study. All 19 participants were UK businesses, mainly engaged in the production of horticultural crops in the UK. The businesses reported that they employ about 1,700 staff full time and about 9,500 seasonal workers.

\begin{figure}[h!]
\begin{center}
\includegraphics[width=0.5\textwidth]{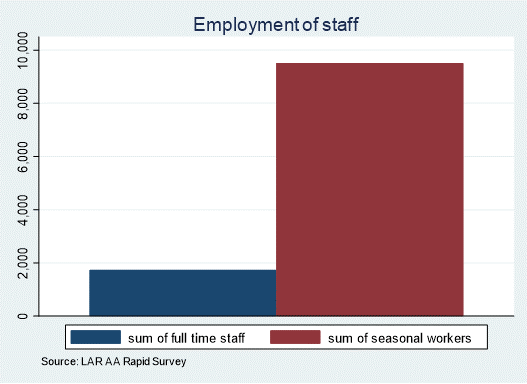}
\caption{Employment of staff}
\label{fig:staff-employment}
\end{center}
\end{figure}

As shown in Figure~\ref{fig:staff-employment}, the total number of seasonal workers represented by these 19 businesses is a considerable share of the overall employment of seasonal workers in UK agriculture: about 15\% of the national total for agriculture and 24\% of the total for the fruit sector only (fruit is $\approx40$k of $\approx64$k total)~\cite{ons_labour_2018}.
The recruitment of seasonal workers has been dominated by migrants for several decades and participants reported using multiple channels to recruit seasonal workers. Most use a combination of recruitment agencies (14 out of 19) and direct recruitment (13 out of 19). Seven out of 19 specifically mentioned using the Seasonal Agricultural Workers Scheme (SAWS). Only a few (5 out of 19) mentioned engaging UK citizens for seasonal labour, with one recruiting UK workers nationally/online and the others more on a `local' basis.

\subsection{Labour availability}
In this section participants' information about current (2020) labour availability and the implications it has for the future (2021) production planning was analysed. The main purpose was to investigate current labour challenges and potential implications for the 2021 growing season.

\subsubsection{Impact of Covid-19 and/or Brexit during 2020 on the availability of labour}

Impacts of Covid-19 and/or Brexit during 2020 were mainly related to labour shortages, particularly for those who had not anticipated the magnitude of the challenge they would face and who had to make do with the available labourers. This had a direct impact on output and increased the costs of production, i.e. training cost and low productivity from workers who were not used to this kind of farm work.  The majority of participants (14 out of 19, or 74\%) reported moderate or severe labour shortages during 2020. The sector was faced with dramatic reduction in the number of available labourers, high turnover, and a reduction in the number of returnees. Some of the common responses were:
\begin{itemize}
\item \emph{`We only received half of our returnee numbers in the spring$\ldots$'}
\item \emph{`Reduced dramatically'}
\item \emph{`Significant shortages'}
\item \emph{`Higher turnover'}
\item \emph{`we tried to employ British staff this season and for us it did not work due to poor production and work ethic.'}
\end{itemize}

While labour shortages and associated factors were observed by a majority of the businesses, 5 out of 19 participants reported that their labour was less affected by the impact of Covid-19 and Brexit because of the existing labour plans the business had in place.  In general, Covid-19 impacts were clearer in the responses than Brexit-related ones, with a more severe impact apparent for participants needing or securing labour later in the year. Reported impacts largely included travel disruptions and general worker incentives. For example:
\begin{itemize}
\item \emph{`Both of these events have created a large amount of uncertainty and fear amongst workers from abroad which has reduced the amount of people available and also increased unplanned/early departures from the farm leaving us short staffed.'}
\end{itemize}

Brexit-related impacts were mainly discussed as emergent, contributing to wider 2020 labour availability issues dominated by Covid-19 impacts. However, it was apparent from responses that some participants had already undertaken substantial preparatory work for the end of the Brexit Transition period.

Two participants cited efforts in 2020 to maximise EU worker settlement in the UK to mitigate impacts. However, others mentioned concerns that they have been unable to exploit the settlement scheme in this way because workers have not signed up for the required status or because the exchange rate is not attractive to overseas workers. Although the question was explicitly limited to 2020, Brexit tended to be treated as a prospective impact in participants' responses.  However, it should be noted that the survey was conducted before the UK and EU Deal was negotiated, when there was considerable uncertainty as to whether a deal would be concluded and, even if it was, what it would contain.
Many responses regarded the potential longer term impacts from Brexit as more significant than those felt in 2020 due to Covid-19. For example:
\begin{itemize}
\item \emph{`Very significant concerns about labour availability for 2021.'}
\item \emph{`For 2021 Brexit will be major problem in absence of SAWS scheme.'}
\end{itemize}

\subsubsection{Impact of labour challenges on 2021 production plans and growing season}

A clear majority of the respondents (14 out of 19, or 74\%) predicted that the labour challenges experienced in the industry will have an impact on their 2021 production plans, while the remaining 26\%, were still unsure if labour supply issues would have any impact on their production plans. Particularly interesting was that some businesses said that plans had not been affected because ‘we do not know how to change’ as opposed to any lack of need to change.

The dominant impact on production plans in 2021 was reported to be scaling back production.  Other cited actions were: increasing efforts to secure new sources of workforce; accelerating automation; and, over a third were adopting a wait and see approach (or had no plans to change). In terms of scaling back production, most responses indicated that absolute production levels are expected to fall in 2021.

\begin{table}[h!]
\begin{center}
\caption{Impact on 2021 plans}
\label{tab:impact-on-2021}
\begin{tabular}{|c|c|c|c|c|}
\hline
\emph{Response} & \emph{Scale back} & \emph{Increase labour} & \emph{Accelerate} & \emph{Unknown / wait and see} \\
& \emph{production} & \emph{related efforts} & \emph{automation} & \emph{(or no plans to change} \\
& & & & \emph{-- see text)} \\
\hline
\# (\%) & 9 (47\%) & 4 (21\%) & 2 (11\%) & 7 (37\%) \\
\hline
\end{tabular}
\end{center}
\end{table}

As indicated in Table~\ref{tab:impact-on-2021}, three of the participants included in the `unknown / wait and see' category simply reported no plans to change and provided no further information; and a further participant's responses simply indicated general uncertainty without providing much further detail. Responses showed that businesses felt that greater automation and reduced labour reliance would be the right direction for longer term production plans, but many felt that this policy is currently impossible to deliver as a result:
\begin{itemize}
\item \emph{`[We need to] reduce [our] need [for] staff by [using] more machines… [but there is] no security from customers to make those investments.'}
\end{itemize}
\noindent
Sensitivity to the need to appeal to local British labour pools and to draw on them in competitive ways (including through increased automation) was also apparent: 
\begin{itemize}
\item \emph{`[local] pool of people so we want to be most attractive (pay and conditions) for available labour'}
\item \emph{`in heavy competition with Amazon, losing staff to them and so had to put [wage] rates up'.}
\end{itemize}
Responses which focused on plans to increase labour supply included reflections on the need for government pilot schemes to continue, to use labour supply agencies or to adopt new strategies to recruit staff. They also reflected on the need for managerial challenges associated to language skills, work ethics and Covid-19 social distancing requirements. Only 2 participants gave answers which mentioned automation as a key action for 2021.

\subsection{Productivity in the workforce}

In this section, participants responded about workforce productivity trends and the wider impacts of productivity i.e. on costs or quality of work (e.g. due to less skilled workers). 74\% of participants (14 of the businesses) had seen changes in the productivity of their workforce in the 2020 growing season, while 26\% (5) had not seen any changes.

Of the fourteen respondents who had seen changes in productivity: two participants answered exclusively in terms of decreased worker quality and efficiency; two answered exclusively in terms of increased costs – one estimated the increase of 8\% and the other claiming a year-on-year labour cost increase of about 20\%; with ten participants observing both changes in terms of quality and cost. For example:
\begin{itemize}
\item \emph{`Our inability to access enough skilled workers and having to ``make do'' with a proportion of lower/no experience level people$\ldots$ has reduced the productivity of the farm by increasing costs both directly (the worker is less productive) and indirectly (the worker requires much more supervision and training) as we strive to maintain our standards.'}
\item \emph{`[In] quarantine teams [it is] difficult to drive productivity. People movement costs up significantly. Costs up. Quality of work down.'}
\end{itemize}
Participants explained quality and efficiency challenges in different terms, for example social-distancing regulations have had a differential impact on field and packing work while the most pronounced trend was towards reports of lower availability of experienced workers and reduced work ethic:
\begin{itemize}
\item \emph{`Of the less experienced workers: productivity is 40\% lower than experienced workers'}
\item \emph{`Ongoing drop in motivation.'}
\item \emph{`The drive is less noticeable.'}
\item \emph{`The main [bulk] of pickers has been less productive and keen to go for the piece work rates that [enable] them to earn more than basic hourly rate, when picking.'}
\end{itemize}
\noindent
Whilst some responses attributed these observations to the increasing employment of UK nationals, others clearly concerned changes in the profile of non-UK workers:
\begin{itemize}
\item \emph{`Employment of UK nationals was problematic. Training costs, make up pay and raised piece work rates.'}
\item \emph{`Getting people to take on more responsible tasks is proving problematical due to various reasons, which include, lack of language, understanding, length of time here, i.e. is it worth spending on training courses for staff that may not be back next season.'}
\end{itemize}
\noindent
One participant also reported that an increase in worker average age has been relevant to fitness. Another participant estimated productivity falls of 10-20\% depending on type of work. Eleven of the participants reported that changes in productivity impacted overall competitiveness in the marketplace while four responded no impact, and one participant responded that they did not know. Six participants reported an impact on their profitability, one in terms of reduced margins, two in terms of the elimination of profit and three in terms of outright loss. Another response cited an impact on investment and growth:
\begin{itemize}
\item \emph{`Quite simply our increased cost of production has reduced/eliminated any net profit that the business relies upon to reinvest and grow'}
\end{itemize}
Three participants mentioned lack of higher prices as a factor in this context, one reporting some success in forming a cooperative preventing undercutting by other producers but the others simply stating:
\begin{itemize}
\item \emph{`Higher prices are hard to achieve.'}
\item \emph{`our customers, don’t want to put prices up, as result of covid'}
\end{itemize}

\subsection{Growth Potential of the sector}

In this section participants responded about their expectations on the future demand for UK production of their crops and options to meet this demand in relation to the anticipated workload in the short (2021) and longer term (2022 onwards). 

\subsubsection{Expectations of future demand for UK crops}

A substantial majority of participants expected that demand will increase. Fifteen participants expected an increase, while three responded Don’t know and none expected a decrease. In terms of participants expectation to changes in demand, both in the short (2021) or the longer term. Participants expected three aspects of demand to change: (i) quantification of the change, (ii) direct causes of the change, and (iii) interaction between demand and production. Some participants used only one aspect, but others include all three. 

The quantification of the change: participants justified their expectation on the increase of demand based on their 2020 experience when demand was boosted by 2-3\% on vegetables, 20\% on fruits and 40\% on flowers. Based on these results, they were expecting that future demand will increase by 3\% in fruits and 30\% on flowers. This is in line with news reports on the increase in demand for fruit in supermarkets/retailers~\cite{bbc_coronavirus_2020} and, on flowers on-line~\cite{may_seo_2020} during the stay-at-home order. 

Direct causes of the change - participants reported three reasons for the increase in the demand for domestic production: Changes towards a healthier diet and lifestyle during and because of the pandemic was the main cause referred to by seven fruit producers. Three participants, all from the flower sector, also cited limited EU imports due to Brexit/Covid-19 restrictions and high exchange rates as drivers of increased demand. Two producers commented on cooking at home as a factor that increased the demand for fruit and root vegetables. Finally, one participant considered that their customers’ strategy to reduce risks in the supply chain made them adopt procurement policies which are more reliant on domestic production. 

Respondents also highlighted that increased year round demand was also encouraging changes in production systems; like extending the growing season by using tunnels in soft fruits production and thus reducing imported volumes.

This complex interaction between multiple sources of supply and demand was also important to describe the influence of supermarkets in affecting production by offering imported fruits during the British fresh fruit season. Three participants from the fruit and one from the vegetable sector used this type of interaction in the supply chain to explain that although they expect an increase in demand for UK production, they will not expand their production because of a lack of clarity about labour availability and the future of a seasonal labour permit scheme.

For companies who adopt this position, their strategy to tackle the problem of labour shortage is by maintaining or reducing the cultivated area, rather than expanding, and/or by switching to less labour-intensive crops. Participants, especially the ones who responded that they do not know what changes will occur in future UK demand for their crops, cited general economic recession or the negative impact of higher production costs (implicitly felt to imply higher consumer prices) as key factors in determining their position. 

\subsubsection{Short, medium and long term solutions to meet the expected increase in workload}

\emph{In the short term (i.e. 2021),} participants selected the main options they would adopt to meet increased demand from amongst the following options: Changes in labour schemes and policies; Automation; Changes in crop/farm management; and Clients' (ie retailers’) commitment.  Some respondents contemplated using several options together and others focused only on one.  The results are summarised in Table~\ref{tab:industry_options}.

Changes in labour schemes and policies were the most cited short-term option (58\%). Comments on how to address this topic included: Six participants referred to the need to have clarity in policies 
\emph{`that allows the 70,000 seasonal workers to travel into and out of the UK'} and asked for a Seasonal Agricultural Workers Scheme (SAWS) enabling the continued recruitment of EU and non-EU seasonal workers.  New labour schemes with more stable access to reliable and skilled labour by recruiting more local workers and transitioning from seasonal to permanent jobs were discussed by four participants. Even with the popularity of this solution, two participants from the fruit sector expressed their concerns about recruiting local workers due to their lower productivity and poorer performance in comparison to migrant seasonal workers.

Automation appeared as the second most preferred short-term option for meeting increased short-term workloads, selected by 32\% of respondents. Further comments included the following:
one participant referred to 
\emph{`existing automation or new technologies that are sufficiently developed now i.e. autonomous vehicles to fetch and carry, thus reducing the fatigue on workers'.}
Another described the gradual transition to mechanisation depends on farmers' investment potential as technology becomes cost-effective and decreases the labour requirement. For another interviewee automation is the only option as he does not see \emph{`staff as an option'.} Similar to labour supply actions, three respondents considered that automation was not an option because:
\begin{itemize}
\item \emph{`Solutions are not available or unaffordable [for the flower sector].'}
\item \emph{`High specifications [on soft fruit tasks that] can only be obtained by manual picking.'}
\item \emph{`Our [vegetable] sector will take a long time to embrace technology and drive labour requirements down.'}
\end{itemize}

Despite concerns about the ability to adopt automation solutions in the short term (i.e. 2021) and a focus instead on labour market solutions, \emph{in the medium to long term (i.e. 2022 onwards),} participants considered that these strategies needed to be swapped. Virtually all participants saw automation as a key medium to long term solution to increased workload, selected by 17 out of 19 (89\%), whereas only 3 out of 19 (16\%) thought that labour market solutions were still a main option in the longer term.
Results are tallied in Table~\ref{tab:industry_options}.

\begin{table}[h!]
\begin{center}
\caption{Summary of responses: short vs. medium and long term options for industry to meet increased demand.}
\label{tab:industry_options}
\begin{tabular}{|c|c|c|}
\hline
\emph{Options} & 
\emph{Short term (2021)} &
\emph{Medium to Long term} \\
& & \emph{(2022 onwards)} \\
\hline
Automation & 6 (32\%) & 17 (89\%) \\
Farm/crop management & 4 (21\%) & 1 (5\%) \\
Labour supply schemes & 11 (58\%) & 3 (16\%) \\
Clients' commitment & 1 (5\%) & 2 (11\%) \\
\hline
\end{tabular}
\end{center}
\end{table}

Automation was the dominant response for long term options to meet the workload demand.
Six of the seventeen participants who responded that automation was a major option did not elaborate their answers, but answers from the rest of the respondents focused on different aspects of automation and its relationship with labour.
Six of them gave details on the production processes, tasks and methods that can be automated.
Another five interviewees highlighted how automation will become the solution for the labour shortage, reduce the reliance on seasonal workers, respond to the scarcity of skilled and capable workers and, decrease labour costs.
Furthermore, two interviewees argued that automation will need to be aligned with changes in labour schemes such as upskilling current workers and ensuring the supply of staff who can operate the equipment effectively.
Three participants discussed how important it is that technology becomes \emph{`more cost-effective'} to reduce input costs (fertiliser, chemicals, and labour) and to contribute to environmental Key Performance Indicators (KPIs). Two also highlighted the need for a commitment from retailers to establish longer-term working and more collaborative relations with suppliers to give certainty and support investment, as profit margins are too small to make big investments unless they have a secure long term market. One respondent emphasised the role of crop/farm management changes by \emph{`focusing more on cut flower types with lower labour requirement'}. Two participants did not suggest automation as a long-term option for labour shortage: one did not elaborate on their reasons; one focused on labour saying \emph{`We would be given a license to recruit our own workers'.}

\subsubsection{Potential of automation as a solution for workload demand}

Participants' responses concerning the types of jobs to automate referred both to the generic type of job function (e.g. handling) and/or the crop management practices/tasks these were applied to (e.g. weeding, harvesting etc). Crop tasks were the most referred to followed by joint explanations using type of job function and crop tasks. Nine participants reported types of job functions like repetition, transportation, simplification of tasks, and their contribution to improving efficiency and process flow as being important drivers for automation. Transportation was the most cited (6) with phrases like:
\begin{itemize}
\item \emph{`autonomous transport'},
\item \emph{`movement of trays/pallets, etc.'},
\item \emph{`fruit movement from picker to end of field'},
and
\item \emph{`moving fruit from field to coldstore'}.
\end{itemize}

All but 3 respondents talked about both pre- and post-harvest practices such as: planting, crop protection (plant cleaning/weeding), husbandry, pruning, picking-harvesting, processing and, packing and packhouse as important areas for automation, showing that there is demand for automation at every stage of the production cycle. They also cited fruit retrieval, crop scouting and flower grading. The most cited crop tasks were:
\begin{itemize}
\item Picking-harvesting cited by eleven participants; 
\item Packing and packhouse cited by six respondents;
\item Crop protection (plant cleaning and weeding) cited by four interviewees.
\end{itemize}
\noindent
Participants tended to include in their responses two or more crop tasks. The study received complementary responses associated with the use of data science and machine learning for yield prediction and labour recording. Some participants consider that other factors will be key considerations in the automation journey, including: 
\begin{itemize}
\item Growers and worker capacity development; 
\item Technology multitask attributes; and
\item Crop prioritisation.
\end{itemize}
\noindent
The Integration of technology attributes and crop prioritisation were concerns expressed by farmers with particular crops like cut flowers or root vegetables. There were some participants who expressed a clear integrated, or systems based, approach to how robots and autonomous systems can offer improved decision making across a range of crop tasks.

\section{Discussion and Conclusion}
\label{sec:discussion}

The overall indication from this rapid survey is that British horticultural output is at risk of falling substantially in 2021 unless solutions to labour supply challenges can be found.  The industry welcomed the December 2020 announcement by government that the SAWS programme would be expanded to 30,000 workers in 2021.  However, there are concerns that even this limit will fail to meet labour demand, there are concerns that flower production has been excluded from the 2021 scheme and considerable uncertainty about what the plans are for SAWS from 2022.

The study concludes that production in the fresh produce and horticulture sector is currently profoundly affected by a combination of Covid-19 and Brexit-related impacts. For producers the short-term uncertainty surrounding these impacts appears to outweigh the longer-term expectation that demand will grow, which is leading to a reluctance to invest and decisions being taken to reduce production in the short term to reduce risk.

\emph{The primary novel result from this study is an indication that UK horticultural production may contract in 2021 to a far greater degree than expected.} Nearly half (47\%) of respondents reported plans to scale back production in 2021 due to labour availability challenges. A further third (37\%) reported that it is too early to decide given the uncertainties they face. More than two-thirds (69\%) reported negative competitiveness impacts due to changes in labour supply. Other results tended to conform with observations from secondary sources, indicating that this sector is experiencing; severe problems with labour availability; and a combination of associated labour cost and quality issues~\cite{bradley_quantitative_nodate,duong_review_2020}. 

Overall, the results of this study suggest that chronic short-term uncertainty is preventing UK horticultural producers from exploiting Covid-19 (e.g. the demand for healthier food) and Brexit (e.g. potential increases in the demand for UK products) impacts as opportunities. They appear to be unable to plan or to justify investment due to labour supply constraints in the context of persistent supply chain precariousness, despite confidence about general demand growth in the sector and the potential of automation as the means to address that growth over the longer term.

Our discussion focuses on sectoral productivity through automation, grounding the challenge of global competitiveness in more localised labour issues. These results suggest a sector obliged to compete over a lower-quality and higher-cost labour supply. Rather than one that is supported to innovate in ways that reduces reliance on labour, develops better-quality jobs and delivers on the global objective of making supply chains more resilient to future disruptions~\cite{aday_impacts_2020, henry_innovations_2020}.

A contraction in the scale of UK fresh produce would also restrict the domestic supply of products, which can deliver environmental benefits (UK production is typically lower carbon and lower water use) and address public health needs for dietary change. 
UK consumer preferences are also increasingly focusing on the UK local and regional supplies. However, if productivity goes down, imports will go up (along with GHG cost) and food prices (whether UK produced or imported); a regressive tax on society, especially low-income individuals, and moves them from eating healthy food (``fresh produce''). The exact opposite of all UK government policy on many fronts and counter argument to Dimbleby's Part one of the National Food Strategy~\cite{dimbleby_2020}, particularly the need to address public health needs for dietary change.

Closing observations focus on the policy recommendations suggested by this study.  It is clear that growers favour a policy which combines short term labour supply commitments with the acceleration of medium and long term investments in automation, aligned with UK Agri-Tech Strategy~\cite{defra_uk_2013} and the Industrial Strategy.  A proactive automation agenda would both meet growers' workload needs at the same time as delivering higher quality jobs which are more attractive to UK workers.

The study also suggests additional discussion points for the research community: methodological standards for timely, policy-oriented research on the role of short-term effects as determinants of economic outcomes are needed because of the increased speed of response which is needed.  Whilst Industry 4.0 and similar changes in the workforce were already leading to this need, Covid-19 and Brexit Transition have both accelerated the speed of change, across all industries, driving a need for more responsive policy making.

\section*{Acknowledgments}
This study was supported by the \emph{Lincoln Agri-Robotics} grant, funded by  UKRI Research England under the Expanding Excellence in England programme.

\nocite{*}
\bibliographystyle{apalike}
\bibliography{AA}

\end{document}